\documentclass[journal, 11pt]{IEEEtran}
\usepackage{graphicx}
\usepackage{float,lscape}
\usepackage{epstopdf}
\usepackage[hyphens]{url}
\usepackage{url}
\usepackage{algorithmic}
\usepackage{xr}
\usepackage{booktabs}
\usepackage{longtable}
\usepackage{tabularx}
\usepackage{graphicx}
\usepackage[table,xcdraw]{xcolor}
\usepackage{amsmath}

\usepackage{multirow}
\usepackage{balance}
\usepackage{algorithmic}
\usepackage[table,xcdraw]{xcolor}
\usepackage[normalem]{ulem}
\useunder{\uline}{\ul}{}
\usepackage[]{algorithm2e}
\usepackage{ragged2e}
\usepackage{enumitem}
\usepackage{balance}
\usepackage{xcolor}
\usepackage[switch]{lineno} 
\usepackage{stfloats}

\usepackage{tikz,xcolor,hyperref}

\definecolor{lime}{HTML}{A6CE39}
\DeclareRobustCommand{\orcidicon}{%
	\begin{tikzpicture}
	\draw[lime, fill=lime] (0,0) 
	circle [radius=0.16] 
	node[white] {{\fontfamily{qag}\selectfont \tiny ID}};
	\draw[white, fill=white] (-0.0625,0.095) 
	circle [radius=0.007];
	\end{tikzpicture}
	\hspace{-2mm}
}

\foreach \x in {A, ..., Z}{%
	\expandafter\xdef\csname orcid\x\endcsname{\noexpand\href{https://orcid.org/\csname orcidauthor\x\endcsname}{\noexpand\orcidicon}}
}

\ifCLASSOPTIONcompsoc
  \usepackage[nocompress]{cite}
\else
  \usepackage{cite}
\fi

\ifCLASSINFOpdf
\else
\fi
\begin{document}

\runningpagewiselinenumbers
% \linenumbers

\title{Blockchain-based Collaborated Federated Learning for Improved Security, Privacy and Reliability} 

% \title{Split-step Wavelet Transform Computational Algorithm to Estimate Path-loss Over Rough Sea Surface}

\author{Amir Afaq$^{1}$, Zeeshan Ahmed$^{1}$\orcidA{},Noman Haider$^{2}$\orcidB{}, Muhammad Imran$^{1}$\orcidC{}

\begin{footnotesize}
$^{1}$ School of Engineering, Information Technology and Physical Sciences, Federation University, Brisbane, Australia. \\

$^{2}$ College of Engineering and Science, Victoria University, Sydney, NSW, 2000, Australia.\\ 
Email: \{zeeshanahmed, noman90, dr.m.imran\}@ieee.org

\end{footnotesize}}
%$^{1}$, Komal S. Khan$^{1}$ \\
% \begin{footnotesize}
% $^{1}$ College of Science and Engineering, James Cook University, Cairns 4870, Australia.\\
% \\
% %Corresponding author: nhuss@usn.no
% \end{footnotesize}}

\vspace{-2.0cm}

\markboth{Draft}
{}

\IEEEtitleabstractindextext{
\begin{abstract}\justifying 
Federated Learning (FL) provides privacy preservation by allowing the model training at edge devices without the need of sending the data from edge to centralized server. FL has distributed the implementation of ML. Another variant of FL which is well suited for the Internet of Things (IoT) is known as Collaborated Federated Learning (CFL), which does not require an edge device to have a direct link to the model aggregator. Instead, the devices can connect to the central model aggregator via other devices using them as relays. Although, FL and CFL protect the privacy of edge devices but raises security challenges for a centralized server that performs model aggregation. The centralized server is prone to malfunction, backdoor attacks, model corruption, adversarial attacks and external attacks. Moreover, edge device to centralized server data exchange is not required in FL and CFL, but model parameters are sent from the model aggregator (global model) to edge devices (local model), which is still prone to cyber-attacks. These security and privacy concerns can be potentially addressed by Blockchain technology. The blockchain is a decentralized and consensus-based chain where devices can share consensus ledgers with increased reliability and security, thus significantly reducing the cyberattacks on an exchange of information. In this work, we will investigate the efficacy of blockchain-based decentralized exchange of model parameters and relevant information among edge devices and from a centralized server to edge devices. Moreover, we will be conducting the feasibility analysis for blockchain-based CFL models for different application scenarios like the internet of vehicles, and the internet of things. The proposed study aims to improve the security, reliability and privacy preservation by the use of blockchain-powered CFL.
\end{abstract}

}
\maketitle

\IEEEdisplaynontitleabstractindextext
% \IEEEdisplaynontitleabstractindextext has no effect wen using
\IEEEpeerreviewmaketitle
\IEEEpeerreviewmaketitle

\section{Introduction}
\label{introduction}
The recent technological advancements in information and communication technologies have completely transformed every human life, however, this growth has also raised some serious security and privacy challenges. Machine Learning (ML) is envisioned to be a promising solution to address the security and privacy challenges and vulnerabilities in the existing and future digital infrastructure. Moreover, the changing landscape of information and communication systems (ICS) has brought forth unique security, privacy and reliability challenges \cite{tourani2017security}. For instance, because of COVID-19, there has been 600 \% increase in cybercrime where complex phishing attacks have been used and many have fallen victim to it \cite{CybsecStats}. The malware attacks have seen tremendous growth in last ten years along with 812.67 million malware attacks in 2018 alone. Same is the case with the malware attacks which rose 350\% in 2018 worldwide \cite{CybsecStats}. Hence, our ICS are more prone to cyber attacks than ever. Malicious actors also are using machine-powered sophisticated attacks to beat the defense mechanism in place and any such breaches have proven to be catastrophic for the victim network. Moreover, the privacy concerns are also ever-high because of data breaches, theft and data leaks at top-tier digital media companies. Many of such breaches resulted into sever consequences and proven to be fatal as well in some cases like Ashley Madison data breach and cyber attacks on hospitals \cite{lewis2002assessing}.  

\subsection{Machine learning for Security}
The recent advancement in artificial intelligence (AI) and Machine Learning (ML) has the capability protect the digital infrastructure against sophisticated and ever-evolving cyber attacks \cite{truong2020artificial}. Because of robust, dynamic and real-time adaptive nature of ML models, they are well-suited as counter-measures for machine-powered attacks from threat actors. The AI and ML has open gateways in the domains of security, privacy, threat detection for ICS and helped in developing effective methods for the protection of ICS. The concept of using ML and AI in security and privacy is not new but the importance of these ML and AI methods have been highlighted by the evolution of deep learning algorithms. Most of the methods before the development of deep learning were dedicated to model the attack patterns with certain characteristic which are not robust in nature, but with AI and ML, it is expected that systems will become more resilient towards new sophisticated attacks with changing characteristics \cite{calderon2019benefits}.

The ML is showing the positive impact in information security field by adopting ML algorithms to address the security and privacy issues. The information security industry is generating more and more data which open them to advance threats and AI could be a powerful antidote for these threats. The first generation of ML solutions are focused on scrutinizing data, detect threats and assist human in remediation plan. The second generation of ML solution will make the systems more autonomous and leave the critical support issues to humans. There have been different types of ML model implementations based on where the model will be trained and deployed. This seggregation is evolved because of the privacy issues when transferring ML model or its parameters and training data between the central ML model and edge devices. The edge devices can be of different types. For instance, the most resource constrained yet high in number are Internet of Things (IoTs). Similarly, the ML models may also need to communicate with other use cases like Internet of Health Things (IoHTs), Internet of Vehicles (IoVs), Industrial Internet of Things (IIoTs) and Unmanned Air Vehicles (UAVs). Based on how the ML model is implemented and information is communicated between ML controllers and edge devices, different ML implementation types are explained in next section along with privacy and security issues raising from these implementations. 

\begin{figure}[b] \centering
\includegraphics[width=0.5\textwidth,height=08cm]{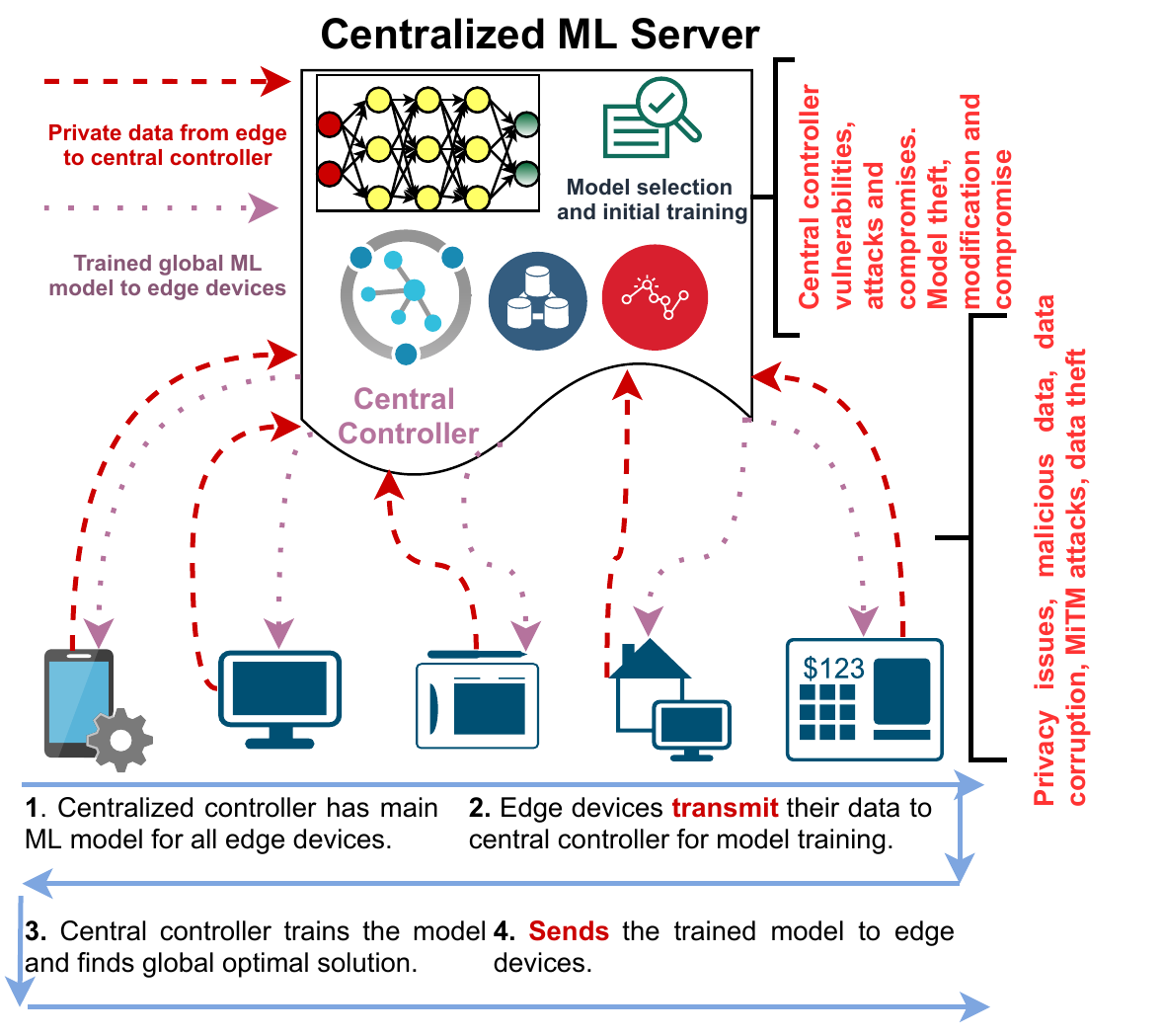}
\caption{Centralized machine learning model implementation from central controller to edge devices and the possible vulnerabilities.}
\label{CML}
\end{figure}

\section{Machine learning Implementation Types}
The ML models can be implemented in different ways. Among these include centralized implementation of global ML models, and distributed implementation of ML known as Federated Learning (FL). There is also Collaborated FL which does not require a direction link between the edge devices and central model aggregator. In this section, we will explain these three types of implementation, security and privacy concerns in these implementations. 
\subsection{Centralized ML}

In this type of ML implementation, the model is selected at centralizer ML server and all the edge devices need to transmit their \textbf{private data} to the central server for ML model training. The central server and controller trains the ML model from the data sent by the edge devices. Once the model is trained and ready of implementation, the central controller than sends the model to the edge devices for implementation. There are multiple security and privacy issues with centralized ML implementation,

\begin{enumerate}
  \item The private data of edge devices can be very sensitive and prone to theft during the transmission. Especially, data-sensitive applications and use cases like IoHTs and IoTs may have serious privacy concerns. 
  \item The trained ML model can be intercepted and modified during the transmission from central server to the edge devices. Such malicious attack may cause severe consequences as the edge devices may not even detect that the ML model has been compromised.
  \item The ML model in itself is pron adversarial type attacks and the compromised data can be fed when intercepted during transmission from edge devices to corrupt the ML model.
  \item The central model could also also be a single point of failure and attacking, risking the entire system.
  \item Another concern for resource-constrained edge devices in IoT, and IoHTs setups is that it will be challenging to transmit their data while having limited battery powers.
\end{enumerate}
Although centralized ML has privacy issues, but it is well-suit for private environments where entire end-to-end communication link is protected. The centralized ML model implementation scenario and exchange of data and model parameters along with their privacy issues is highlighted in Fig. \ref{CML}. More detailed insights on threat surface and attack scope in different parts of ML architecture can be found in \cite{papernot2016towards}. Because of the privacy issues in centralized ML implementation, Google proposed an idea of distributed implementations of ML model known as Federated Learning (FL) \cite{bonawitz2019towards}.

\subsection{Federated Learning}
\begin{figure}[t] \centering
\includegraphics[width=0.5\textwidth, height=08cm]{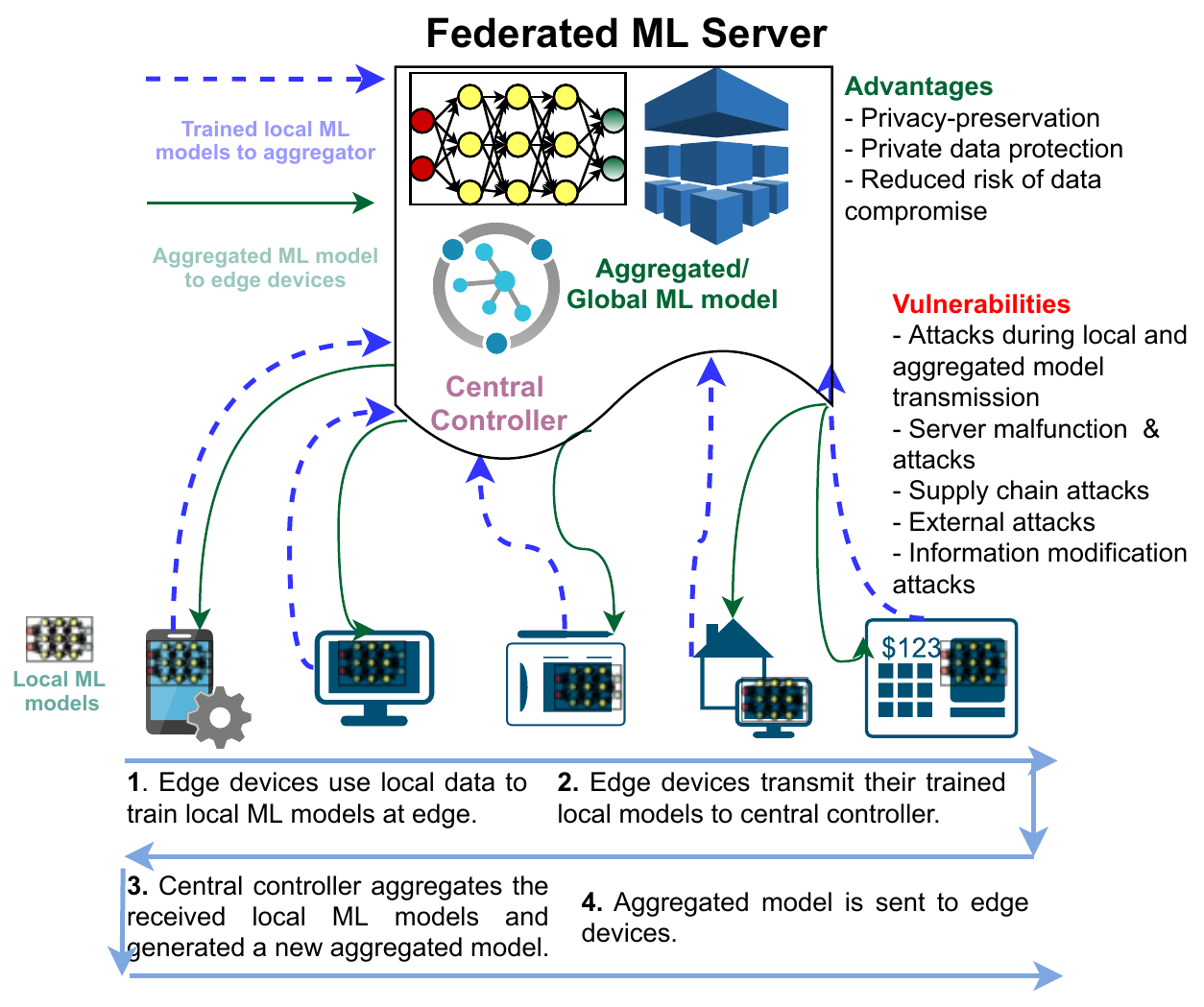}
\caption{The distributed implementation of ML model using FL along with its privacy advantages and risks.}
\label{FL}
\end{figure}

The FL is a distributed implementation of centralized ML. The primary advantage of FL is privacy-preservation for edge devices because edge devices does not have to transmit their private data to central controller. The edge devices have their own local ML model while global ML model resides in central controller. Edge devices use their private data and train their local model and then send their model parameters to the central controller which then develops aggregated ML model (shared optimal solution) from all of the received models from the edge. The aggregated model is finalized at the central controller and model updates along with the model parameters are sent to the edge devices. Edge devices can also request for the updates from the central controller. Hence, this type of distributed ML model implementation preserves the privacy of private data of edge devices as they don't need to transmit it to the central controller \cite{yang2019federated}. Moreover, this also significantly reduces the network bandwidth which is consumed for the transmission of private data from edge to ML server. This reduces the data theft and associated risks for edge devices. The FL architecture along with its advantages and security issues is presented in Fig. \ref{FL}.

\subsection{Collaborated Federated Learning}
\begin{figure}[t] \centering
\includegraphics[width=0.5\textwidth, height=08cm]{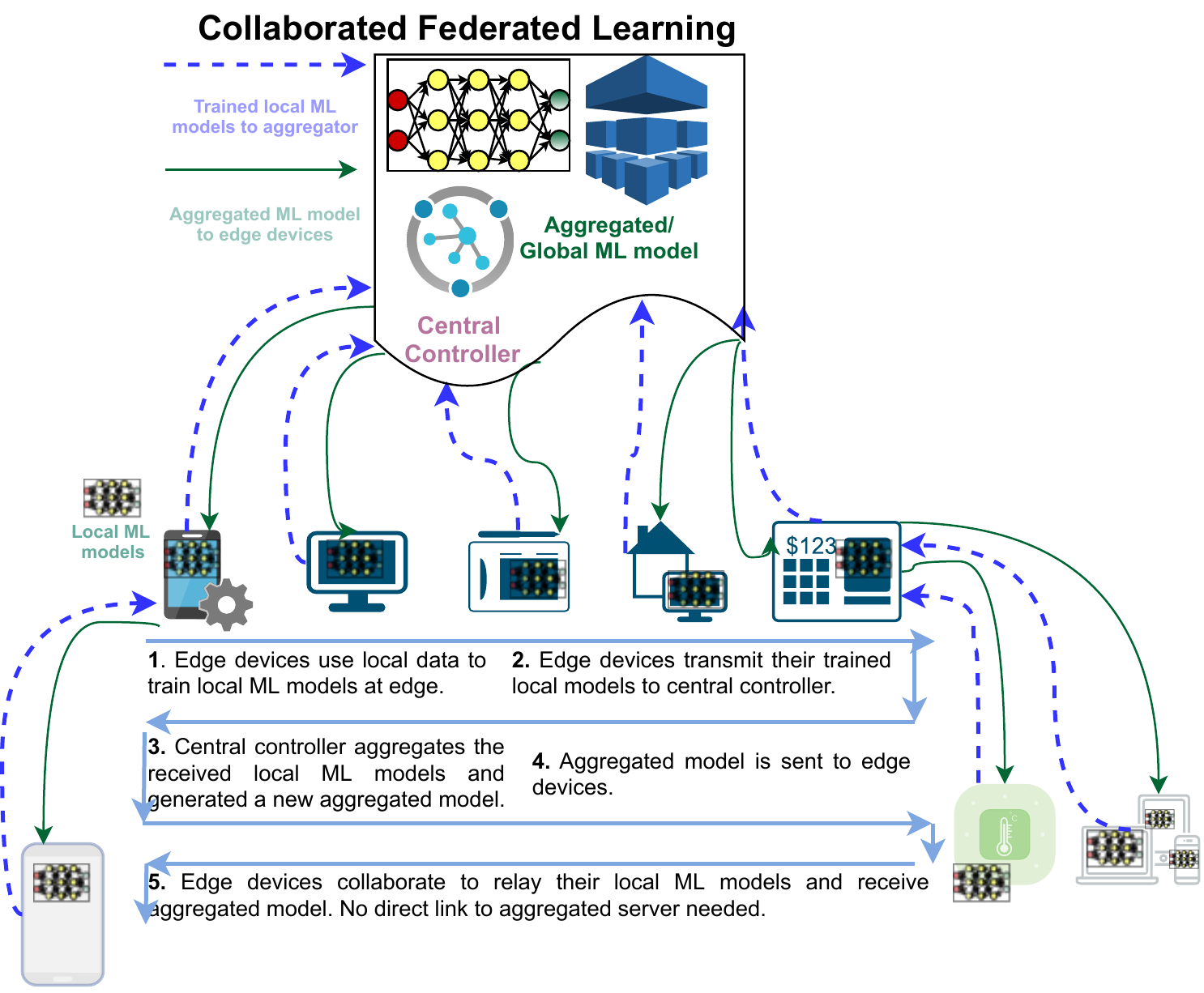}
\caption{The collaborated federated learning model.}
\label{CFL}
\end{figure}

In FL, all the edge devices need to have a link or connection with the central controller. This may not be possible for some devices in remote locations or for the devices relying on the sink or relays for their data transmission. Authors in \cite{chen2020wireless} proposed concept of collaborated FL (CFL) in wireless networks where edge devices can collaborate with one-another and use them as relays to forward their model updates, trained models and communication with the model aggregate. This CFL extends the distributed ML for remote devices which do not have a direct connection with the model aggregator. The CFL has more practical applications and extends the privacy-related benefits of ML by making it scalable. The working mechanism of CFL is presented in Fig. \ref{CFL}.

\subsection{Security and privacy issues in FL and CFL}
The FL and CFL improves the privacy related issues and extends the advantages of ML modelling to the edge. The private data is protected for the edge device as they don't need to transmit it, however the ML model parameters are still being exchanged between model aggregator and edge devices. If this exchange is not secure, consensus-based, and fully secure, then this could still be a major security and privacy risk. The man-in-the-middle could still modify the model exchange parameters, hence making the FL and CFL implementation prone to an external attacks. Hence, there should be a secure, decentralized and consensus-based ML model sharing mechanism which should be trusted by all the devices is needed. 

\begin{figure*}[!htb] \centering
\includegraphics[width=\textwidth, height=08cm]{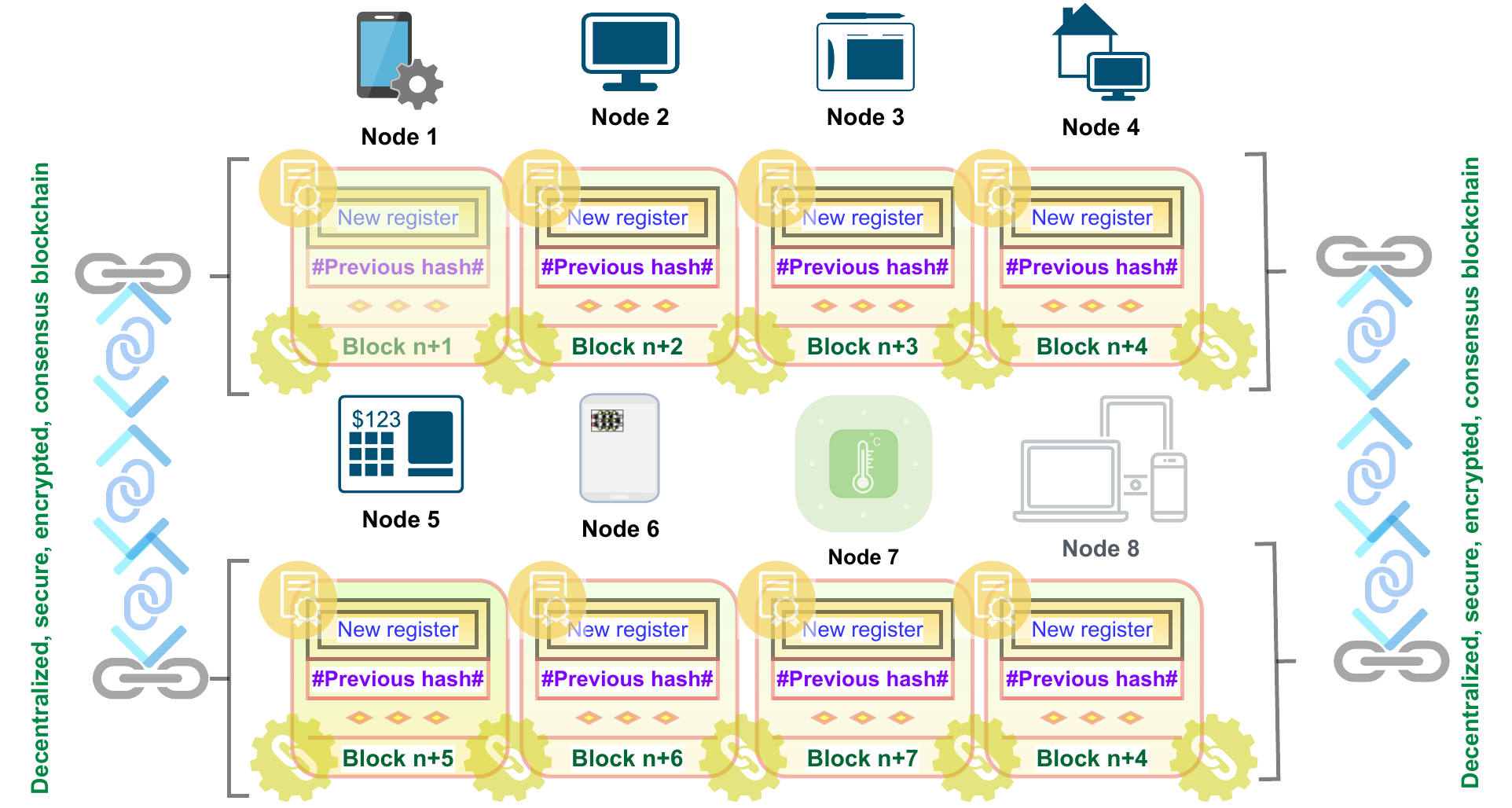}
\caption{Decentralized and distributed blockchain mechanism which uses chain of blocks shared among all the participating devices.}
\label{blockchain}
\end{figure*}

\section{Blockchain}
Blockchain is a decentralized, secure and consensus based information sharing technology which comes with its inherent security and privacy benefits. Devices in blockchain use blocks to add new transactions and share it with other peers using blocks or ledgers, hence all known as distributed ledger technologies (DLT). All the devices in peer-to-peer (P2P) distributed networks are able to verify the integrity of blocks and transactions. Each and every block is securely connected with it's previous and next blocks with strong cryptographic chain which can be verified by all the devices using the blockchain. Hence, making it very secure and hard to tamper with as multiple devices have accurate records of block and tampered blocks can easily be identified and removed by consensus mechanism. The realization of blockchain mechanism is shown in Fig. \ref{blockchain}.

\begin{figure*}[!htb] \centering
\includegraphics[width=\textwidth, height=06cm]{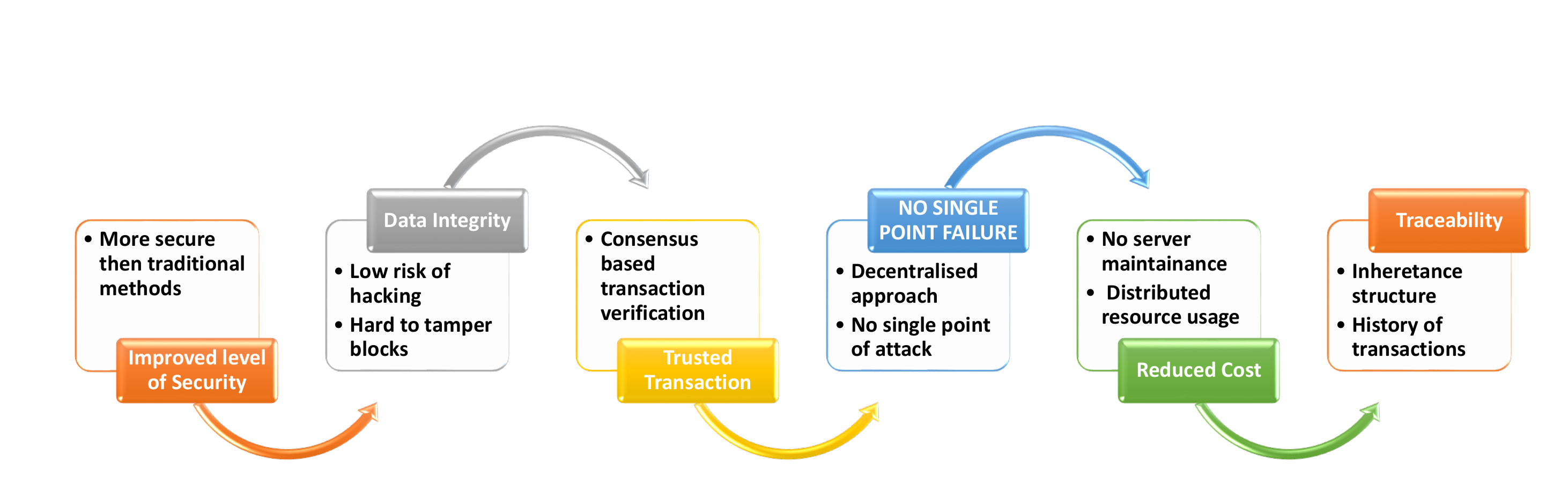}
\caption{Advantages of using blockchain.}
\label{blockchainadvants}
\end{figure*}

Blockchain technology has proven to be potential candidate in privacy-sensitive applications \cite{dorri2017blockchain}. Based on access to certain blockchain network, applications of block chain can mainly be categorized in to public, and private blockchains. Some of the important performance and security advantages of using blockchain are shown in Fig. \ref{blockchainadvants}.

\section{Blockchain powered FL}
The security and privacy challenges related to ML model updates, request and data sharing in FL and CFL can potentially be addressed by using blockchain consensus based mechanism. Devices in private network implementing their local ML models with model updates from global ML model can use blockchain DLT to strengthen the privacy against external attacks. Block chain for FL has been considered a reliable, safe and secure mechanism for devices to share the ML model updates and parameters \cite{lu2019blockchain}. On-device blockchain based federated learning model (BlockFL) has been studied in the literature \cite{kim2019blockchained}. Blockchain powered FL (BlockFL) can also be used for secure information sharing among IoVs \cite{lu2020blockchain, pokhrel2020federated}. 

\section{Research Challenges}
Most of the existing works in the literature have considered BlockFL for privacy-preservation in consensus based data sharing applications. To the best of our knowledge, none of the existing works are using BlockFL for detection, investigation, and analysis of model-poisoning attacks, and backdoor attacks. In this proposed research study, we want to investigate the feasibility of BlockFl for, 
\begin{enumerate}
  \item Detection, investigation and analysis of malicious ML model updates.
  \item Design and modelling of cyber risk assessment based on different types of adversarial, backdoor and model poisoning attacks.
  \item Reliability and latency analysis of using blockchain. 
  \item Customized and on-demand data aggregation and protection strategy.
  \item Different use-cases and applications.
\end{enumerate}

\subsection{Blockchain powered CFL}
Based on the research gap we have identified, we will be investigating the use of blockchain for collaborated FL. To the best of our knowledge, none of the existing works use BlockCFL mechanism for model updates among the edge devices. Considering the scalability of CFL and more wider range of possible application, using blockchain for model updates is expected to open some more future research directions. 

\section{Conclusion and Future Work}
In this work, we have studied the blockchain and federated machine learning for safe, secure and reliable data transfer. Based on our research, we have identified that blockchain can complement the performance advantages of collaborated federated learning and could open new research directions for the future. In the proposed research work, we intend to use block chain for CFL (BlockCFL) and study it's performance advantages in the context of security, reliability and privacy. Moreover, in future work, we are also working on developing the taxanomy of the existing works and their performance advantages and shortcomings.

% \begin{figure*}[t] \centering
% \includegraphics[width=\textwidth, height=8cm]{images/researchplan.png}
% \caption{Proposed research plan for Ph.D.}
% \label{researchplan}
% \end{figure*}

\bibliographystyle{ieeetr}
\bibliography{paper}

\end{document}